\documentstyle[epsf]{l-aa}

\begin{document}

\thesaurus{6(08.03.4; 08.05.3; 08.09.2 Sk$-$69 271; 08.09.2 Sk$-$69 279; 
08.13.2; 09.02.1)}

\title{Two Ring Nebulae around Blue Supergiants\\ 
in the Large Magellanic Cloud}

\author {K.\ Weis \inst{1,2}\thanks{Visiting Astronomer, Cerro Tololo 
Inter-American Observatory, National Optical Astronomy Observatories, 
operated by the Association of Universities for Research in Astronomy, Inc., 
under contract with the National Science Foundation.} 
\and Y.-H.\ Chu \inst{2}$^{\star}$
\and W.J.\ Duschl \inst{1,3} 
\and D.J.\ Bomans \inst{2}$^{\star,}$\thanks{Feodor-Lynen Fellow of the 
Alexander von Humboldt Foundation}
}

\offprints{K. Weis,\\ 
email: kweis@ita.uni-heidelberg.de}

\institute{
Institut f\"ur Theoretische Astrophysik, Tiergartenstr. 15, D-69121 
Heidelberg, Germany
\and
University of Illinois, Department of Astronomy, 1002 W. Green Street, 
Urbana, IL 61801, USA
\and
Max-Planck-Institut f\"ur Radioastronomie, Auf dem H\"ugel 69, D-53121 Bonn, 
Germany}

\date{received; accepted}

\maketitle

\begin{abstract}

Ring nebulae are often found around massive stars such as
Wolf-Rayet stars, OB and Of stars and Luminous Blue Variables (LBVs).
In this paper we report on two ring nebulae around blue supergiants
in the Large Magellanic Cloud. The star Sk$-$69 279 is classified as 
O9f and is surrounded by a closed shell with a diameter of 4.5\,pc.
Our echelle observations show an expansion velocity of 14\,km\,s$^{-1}$ and a
high [N{\sc ii}]$\lambda$6583\AA/H$\alpha$ ratio. 
This line ratio suggests nitrogen abundance enhancement 
consistent with those seen in ejectas from LBVs.
Thus the ring nebula around Sk$-$69 279 is a circumstellar bubble. 

The star Sk$-$69 271, a B2 supergiant, is surrounded by an H$\alpha$
arc resembling an half shell. 
Echelle observations show a large expanding shell 
with the arc being part of the approaching surface.
The expansion velocity  is $\sim$ 24\,km\,s$^{-1}$ and the
[N{\sc ii}]$\lambda$6583\AA/H$\alpha$ is not much higher than that of the 
background emission. 
The lack of nitrogen abundance anomaly suggests that the expanding shell
is an interstellar bubble with a dynamic age of 2 $\times$ 10$^5$ yr.


\end{abstract}

\section{Introduction}

Massive stars are known to have strong stellar winds and lose a lot of 
mass. For example, stars with an initial mass above 35\,M$_{\sun}$ will 
lose 50\% or more of their mass before their demise (Garc{\'\i}a-Segura 
et al. 1996a, 1996b). These
stars have a large impact on the interstellar environments and influence 
the circumstellar surroundings throughout their evolutionary phases. 
Ring nebulae around massive stars testify the effects of stellar mass loss,
as they are formed by fast stellar wind sweeping up ambient interstellar 
medium, fast wind interacting with previous slow wind, or outburst-like 
ejection of stellar material (Castor et al. 1975;  Weaver et al. 1977; 
Chu 1991).

All massive stars that experience fast stellar winds either currently or
previously, e.g. OB supergiants, ought to be surrounded by ring nebulae.
Surprisingly, only a handful of ring nebulae around O supergiants, but not B
supergiants, are known in our galaxy (Lozinskaya 1982; Chu 1991); no ring
nebulae around single O or B supergiants are known in the Large Magellanic 
Cloud (LMC), which otherwise hosts a large collection of shell nebulae
of all sizes (e.g. Davies et al. 1976).  While the scarcity of known ring 
nebulae around OB supergiants could be caused by the lack of a sensitive 
systematic survey, other causes cannot be excluded.

Recently, we found two ring nebulae around blue supergiants in the LMC: 
a closed shell with a 18$\arcsec$ diameter around the star Sk\,$-$69\,279
(designation from Sanduleak 1969), 
and a half shell of 21$\arcsec$ diameter around the star Sk\,$-$69\,271 
(Weis et al. 1995).  As shown in Fig.\ 1, both stars are located to the 
north-east of the H\,{\sc ii} region N\,160 (designation from Henize 1956).
We have obtained additional images of this field and high-dispersion 
long-slit echelle spectra of these ring nebulae in H$\alpha$ and 
[N~{\sc ii}] lines.  These data allow us to determine not only the 
physical structure of the nebulae, but also diagnostics for N abundance
in the nebulae, which can be used to constrain the evolutionary 
states of the central stars.

This paper reports our analysis of the ring nebulae around Sk\,$-$69\,271 
and Sk\,$-$69\,279.
Section 2 describes the observations and reductions of the data; sections 
3 and 4 describe our findings for Sk\,$-$69\,279 and Sk\,$-$69\,271, 
respectively. We discuss the formation and evolution of these two ring
nebulae in section 5, and conclude in section 6.

\section{Observation and data reduction}

\subsection{Imaging}

We obtained CCD images with the 0.9\,m telescope at Cerro Tololo 
Inter-American Observatory (CTIO) in January 1996.  The 2048$\times$2048 
Tek2K3 CCD used had a pixel size of 0$\farcs$4.  The field of view was
13$\farcm$5 $\times$ 13$\farcm$5, large enough to  encompass
both Sk\,$-$69\,279 and Sk\,$-$69\,271. Broad-band Johnson-Cousins B, V, R 
filters and narrow-band H$\alpha$ and [O{\sc III}] filters were 
used. The H$\alpha$ filter had a central wavelength of 6563\,\AA\ and a 
filter width of 75\,\AA, which included the [N{\sc ii}] lines
at 6548\,\AA\ and 6583\,\AA. The [O{\sc III}] filter had a central 
wavelength of 5007\,\AA\ and a width of 44\,\AA. 
The exposure time was between 10 and 300\,s for the B, V and R filters 
and 900\,s for H$\alpha$ and [O{\sc III}].
The seeing was around 2$\arcsec$ during the observations; the sky
condition was not photometric.

Fig.\,1 displays a 6$\arcmin \times 3\farcm5$ sub-field of the H$\alpha$ 
image to show the ring nebulae around Sk$-$69 279 and Sk$-$69 271 and their
relationship to the H\,{\sc ii} region N\,160.  We have subtracted a 
scaled R frame from the H$\alpha$ image to obtain a continuum-free 
H$\alpha$ frame.  Several stars in the field were used for the scaling.  The 
continuum-subtracted H$\alpha$ images (1$\arcmin \times 1\arcmin$) of the 
two ring nebulae are shown in Fig.\,2a and 2b.  Not all continuum sources 
were removed well and some white parts in the images mark the residuals.
Neither of the ring nebula showed emission in the [O\,{\sc iii}] filter. 

We performed a flux calibration of our H$\alpha$ image using a
photo-electrically calibrated PDS scan of Kennicutt \& Hodge's (1986)
Curtis Schmidt plate, kindly provided to us by Dr.\, R.C. Kennicutt. 
We transfered the calibration by using the fluxes of two compact 
emission knots and three narrow H$\alpha$ filaments near N160 that
were visible in both our CCD images and their Schmidt plate.  The 
different spatial resolutions and the variable background levels 
made the largest contributions to the uncertainty in the calibration.
We estimate that the error in the flux calibration should be much less
than 30\%.

\begin{figure}
\epsfxsize=\hsize
\centerline{\epsffile{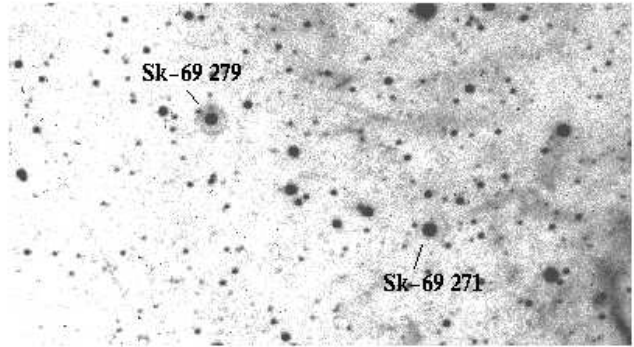}}
\caption{H$\alpha$ image of the region including both ring nebulae. North 
is up, and east is to the left. The field of view is 6$\arcmin \times 3\farcm5$.
The bright H$\alpha$ emission to the west belongs to the H\,{\sc ii} region
N\,160. Sk\,$-$69\,279 is surrounded by a closed shell, and Sk\,$-$69\,271 
a half-shell. 
The straight line near the eastern edge is caused by a satellite
crossing the field of view.}
\end{figure}

\begin{figure}
\epsfxsize=\hsize
\centerline{\epsffile{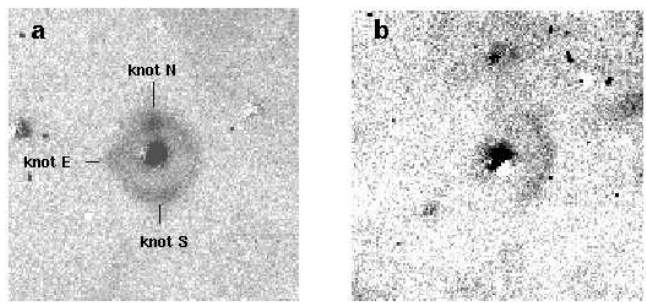}}
\caption{Continuum-subtracted H$\alpha$ images of the ring nebulae around:
(a)  Sk\,$-$69\,279 , (b) Sk\,$-$69\,271.  
North is up, and east is to the left. 
Each image covers 1$\arcmin \times 1\arcmin$.}

\end{figure}

\subsection{Echelle spectroscopy}

To investigate the kinematic structure of the two ring nebulae, we obtained
high-dispersion spectroscopic observations with the echelle spectrograph on 
the 4\,m telescope at CTIO in January 1996.
We used the long-slit mode, inserting a post-slit H$\alpha$ filter 
(6563/75\,\AA) and replacing the cross-disperser with a flat mirror.  A 
79\,l\,mm$^{-1}$
echelle grating was used.  The data were recorded with the long focus red
camera and the 2048$\times$2048 Tek2K4 CCD.  The pixel size was 
0.08\,\AA\,pixel$^{-1}$ along the dispersion and 0$\farcs$26\,pixel$^{-1}$ 
in the spatial axis.  The slit length was effectively limited
by vignetting to  $\sim4^\prime$.  Both H$\alpha$ 6563\,\AA\ and [N\,{\sc 
ii}] 6548\,\AA, 
6583\,\AA\ lines were covered in the setup.  The slit-width was 250\,$\mu$m 
($\widehat{=} 1\farcs 64$) and the instrumental FWHM was about 
14\,km\,s$^{-1}$ at the 
H$\alpha$ line.  The seeing was $\sim 2\arcsec$ during the observations.
Thorium-Argon comparison lamp frames were taken for wavelength calibration
and geometric distortion correction.
The reduction of all images and spectra was done in IRAF.

For the ring around Sk\,$-$69\,279  two east-west oriented slit positions
were observed, one centered on the star itself and the other with a 
4$\arcsec$ offset to the north.  The ring around Sk\,$-$69\,271 was 
observed with only one east-west oriented slit centered on the star.
  
The exposure time was 900\,s for each position. Echelle images of the 
H$\alpha$+[N\,{\sc ii}]$\lambda$6583\,\AA\ lines are presented in Fig.\ 3.
The spectral range shown here is from 6560\,\AA\, to 6600\,\AA; the 
spatial axis is 1$\arcmin$ long in Fig.\ 3a and 3b, and 2$\arcmin$
long in Fig.\ 3c. 
Beside the H$\alpha$ and [N\,{\sc ii}] nebular lines, the geocoronal
H$\alpha$ and three telluric OH lines (Osterbrock et al.\ 1996) can be 
seen (one telluric OH line is blended with the broad nebular H$\alpha$ 
line).  These telluric lines provide convenient references for 
fine-tuning the wavelength calibration.  All velocities given in this
paper are heliocentric.

\begin{figure*}
\epsfxsize=\hsize
\centerline{\epsffile{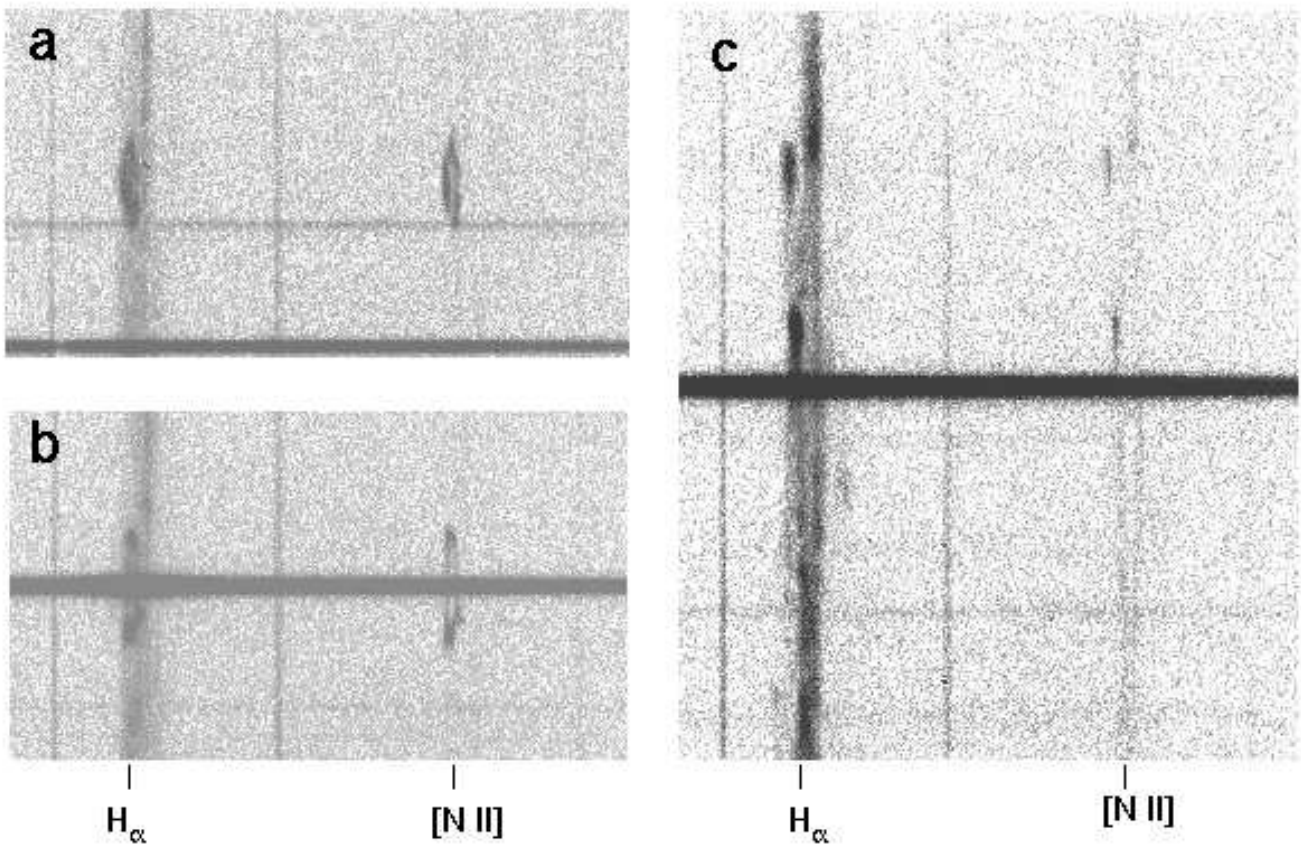}}
\caption{Echellograms of the ring nebulae showing the H$\alpha$ line and the 
[N\,{\sc ii}]$\lambda$6583 line:
(a) centered at 4$\arcsec$ north of Sk\,$-$69\,279,
(b) centered on the star Sk\,$-$69\,279,
(c) centered on Sk\,$-$69\,271.
The spectral axis (horizontal axis) covers from 6560\,\AA\, to 6600\,\AA,
with wavelength increasing to the right.
The spatial axis (vertical axis) is 1$\arcmin$ long in panels a and b, and 
2$\arcmin$ long in panel c. West is up and east is down.}
\end{figure*}

\section{The ring nebula around Sk\,$-$69\,279}
 
Sk\,$-$69 279 is a blue supergiant, as its color and magnitude are
(B-V) = 0\fm05 and V = 12\fm79 (Isserstedt 1975). Its spectral type was
given by Rousseau et al. (1978) as O-B0, and was improved by Conti 
et al.\ (1986) to be O9f.
With $T_{\rm eff} = 30300$ K and $M_{\rm bol} = -9\fm72$ (Thompson et 
al.\ 1982), this star would be located in the very upper part of the HR
diagram (Schaller et al. 1992), making it a massive star, maybe with an 
initial mass larger than 50\,M$_{\sun}$.

\subsection{Structure and Morphology}

As shown in Fig.\,1 and 2a, the nebula around Sk\,$-$69\,279 has a diameter
of 18$\arcsec$.  Adopting a distance of 50 kpc to the LMC (Feast 1991),
this angular size corresponds to a linear diameter of 4.5\,pc. 
Fig.\,1  shows that the nebula is most likely a closed spherical shell
resembling a bubble. Examined closely, the continuum-subtracted image 
(Fig.\,2a) also reveals internal structure of the shell.
There are surface brightness variations along the shell rim;
furthermore, nebula emission extends beyond the shell rim in the north, south
and east directions. We will call these extensions knot N, knot S and knot E,
respectively.
As described in section 3.2, some of these features show kinematic
anomalies as well. 

\subsection{Kinematics of the nebula} 

For Sk$-$69 279 two  long-slit echelle observations were made,
one centered on the star and the other centered at 4$\arcsec$ 
north of the star (Fig.\,3a,b).
Both the ring nebula and the background H\,{\sc ii} region are detected.
The bow-shaped velocity structure originates from the ring nebula around 
Sk$-$69 279 and indicates an expanding shell.
The broad H$\alpha$ component at a constant velocity corresponds to the 
background H\,{\sc ii} region at the outskirts of N\,160. 
Its central velocity at  $v_{\rm hel} \sim 250$ is similar 
to values found through Fabry-Perot measurments by Caulet et al. 1982 
(245.1\,km\,s$^{-1}$) or Ch\'eriguene \& Monnet 1972 (253.3\,km\,s$^{-1}$).
Also the main H\,{\sc i} component in the vicinity of N\,160 is of comparable 
size, at a velocity of  254\,km\,s$^{-1}$ (Rohlfs et al. 1984).  
The velocity profile of the background 
H\,{\sc ii} region with FWHM $\simeq$ 100\,km\,s$^{-1}$, is much broader 
than those typically seen in classical
H\,{\sc ii} regions, indicating a significant amount of turbulent motion. 

To analyze the expansion pattern of the ring nebula we
have made  velocity-position plots, as shown 
in Fig.\,4.
The positions of the data points in the plots are distances 
from the star: the zero-point is the position of the central star, 
negative values are to the east and positive to the west. 
Measurements 
of both H$\alpha$ and [N{\sc ii}] lines are presented for Sk$-$69 279,
but only H$\alpha$ for Sk$-$69 271.
Their error bars are $\pm$4\,km\,s$^{-1}$.
The systemic velocity of the expanding shell, 230\,km\,s$^{-1}$, 
is offset by 20\,km\,s$^{-1}$ with respect to the background 
H\,{\sc ii} region and the H\,{\sc i} gas.

\begin{figure}
\epsfxsize=\hsize
\centerline{\epsffile{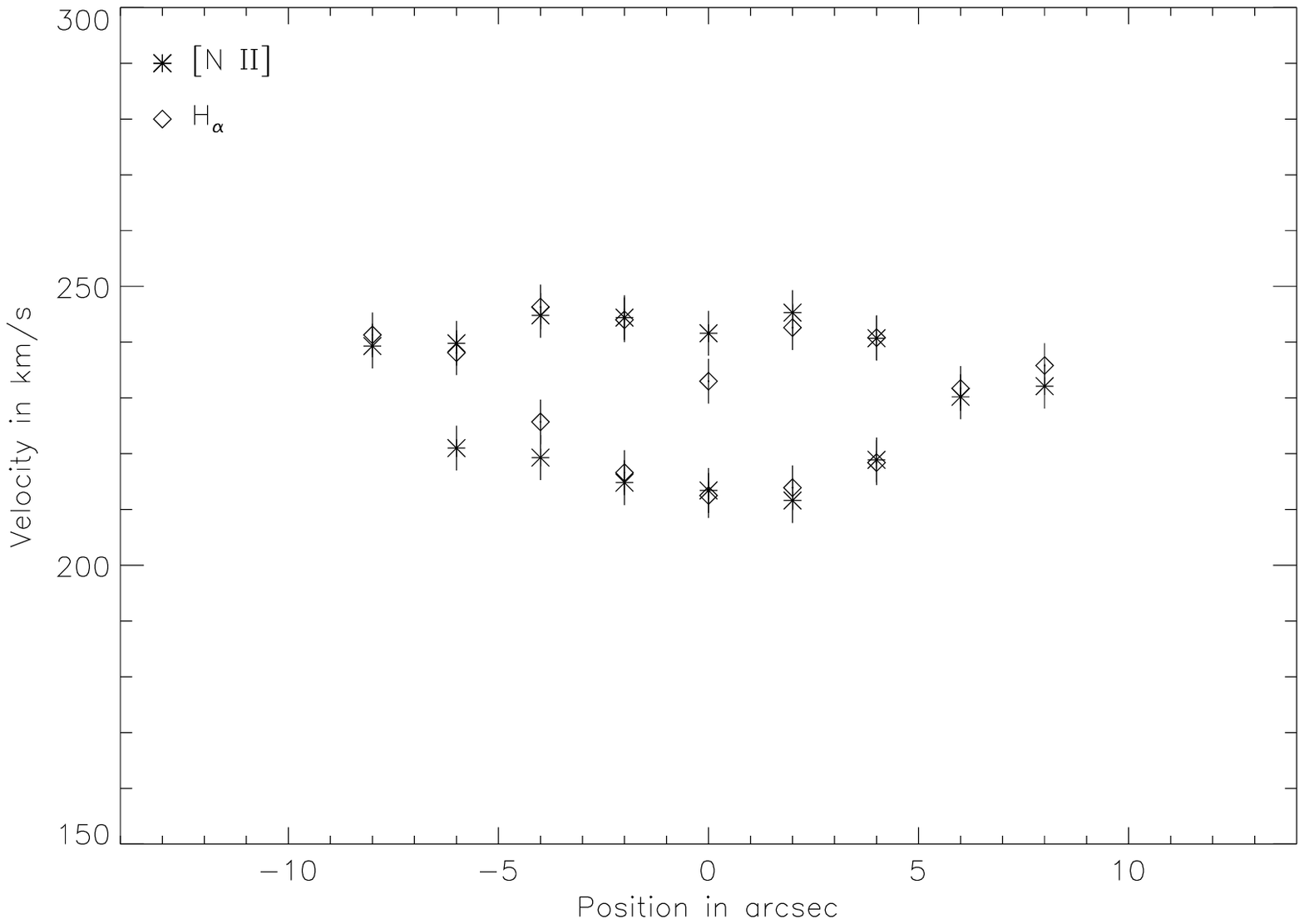}}
\epsfxsize=\hsize
\centerline{\epsffile{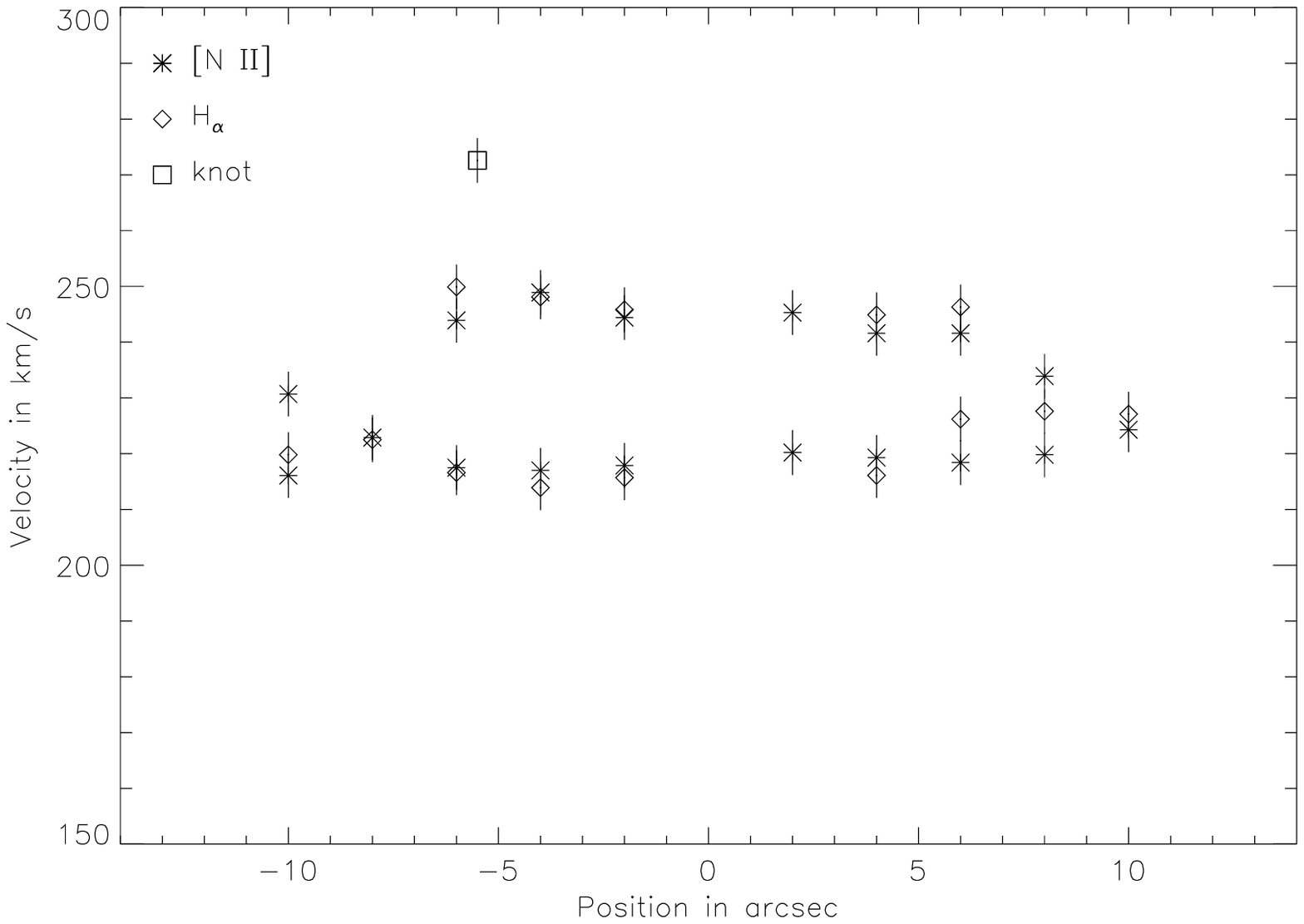}}
\caption{Velocity-position plots for the nebula around Sk$-$69 279.
The upper plot is derived from the echelle slit at 4\arcsec\,
north of the star, and the zero point in the position
axis is 4\arcsec\, due north of the star.
The lower plot is derived from the slit centered on the 
star, and the zero point in the position axis is at the
star. Negative offsets are to the east and positive to 
the west.  Both H$\alpha$ and [N {\sc ii}] line measurements 
are plotted.}
\end{figure}

The position-velocity plot of 
the central slit position (Fig.\,3b) shows typical characteristics of an
expanding shell structure. In Fig.\,3b the approaching side of the shell 
seems to reveal a constant velocity pattern instead of an expansion 
ellipse as seen in Fig.\,3a. This may be explained by a very flat geometry 
of the shell at this position, e.g. nearly no curvature or result in an 
interaction of the shells approaching front with denser interstellar medium
which halts the expansion. 
Despite the noticeable intensity variations in the
shell, the expansion is relatively uniform. The  expansion 
velocity is about 14\,km\,s$^{-1}$ in  both H$\alpha$ and [N\,{\sc ii}] lines
(see Fig. 4).  
However, a knot receding significantly faster than the general expansion 
is detected to the east of the central star (see Fig.\,3b). This knot is  
indicated  with a square in the position-velocity plot (lower plot in Fig.\,4)
and shows a velocity of about 272\,km\,s$^{-1}$.
This knot might be physically associated with the morphologically identified
knot E.

Knot N was partially intercepted by the slit position at 4\arcsec north of the 
central star. It is most likely responsible for the intensity enhancement
on the approaching side of the shell (Fig.\,3a). 
In contrast to knot E, the velocity of knot N does not show noticeable 
deviation from the shell expansion.

Note that the line images of the central slit position (Fig.\,3b) show larger 
intensity enhancement at the shell  rims than those of the slit at 
4\arcsec\, north (Fig.\,3a). This difference in limb brightening is caused 
by the longer path length through the shell at central slit position.

In summary, the ring nebula around Sk$-$69 279 is a closed and uniform 
expanding shell. The morphologically identified knots show different
kinematic characteristics.
Knot N follows the shell expansion, while knot E shows large velocity 
anomaly.

\section{The nebula around Sk$-$69 271}

The color and magnitude of Sk$-$69 271, (B-V) = 0\fm00 and V = 12\fm01 
(Isserstedt 1975) are consistent with those of a blue supergiant in the LMC.  
Using objective prism spectra  Rousseau et al. (1978) classified
the star as B2. This spectral classification may be somewhat uncertain 
because of the low spectral resolution.

\subsection{Structure and the morphology}

Sk$-$69 271 is located in the outskirts of the H\,{\sc ii} region
N\,160 (Fig.\,1). 
The nebula around Sk$-$69 271 consists of only one arc to the west side 
(Fig.\,2b).  It is not clear whether the eastern side is
invisible because it is neutral or is missing because of a real lack of
material. 
Assuming a complete round nebula, the diameter will be 21$\arcsec$, 
or 5.3\,pc. 
Unlike the nebula around Sk$-$69 279, the nebula around Sk$-$69 271 is quite 
uniform and no clumps or knots are discernable.

\subsection{Kinematics of the nebula}

To determine the structure of the nebula around Sk$-$69 271 we obtained 
an east-west orientated long-slit echelle observation centered on the star.
The line image (Fig.\,3c) shows a  complex velocity structure. 
An expanding shell structure 
centered on the star is present and extends over  70$\arcsec$, or 17.7\,pc.
Both the approaching and receding side
are detected with continuous distribution of material.
The average line split near the shell center is about 48\,km\,s$^{-1}$ 
(Fig. 5).
On the edge of the shell the split line converges to the velocity 
of the background H {\sc ii} region at about 248\,km\,s$^{-1}$, marked by the 
dashed line in Fig.\,5.   
The velocity width of the background H{\sc ii} region is 116\,km\,s$^{-1}$.
Superimposed on the expanding shell and the background H{\sc ii} region 
are additional high-velocity components scattering  between 115 and
395 km\,s$^{-1}$.

Interestingly, the H$\alpha$ arc around Sk$-$69 271 corresponds to a brighter
section near the center of the approaching side of the expanding shell,
instead of marking the edge of the expanding shell  (Fig.\,2b and 3c).  
This brighter part appears to lead the expansion on the approaching side.

The velocity structure in the [N\,{\sc ii}] line is similar to that in the
H$\alpha$ line. Unfortunately the  [N\,{\sc ii}] line is too weak for 
accurate velocity measurements. 

\begin{figure}
\epsfxsize=\hsize
\centerline{\epsffile{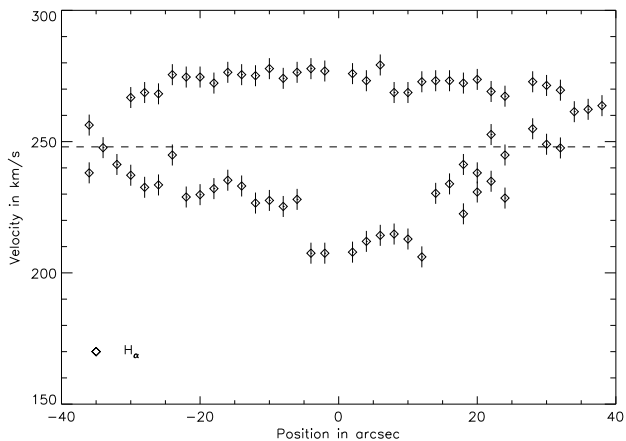}}
\caption{Velocity-position plots for the nebula around Sk$-$69 271.
Only H$\alpha$ measurements are plotted. 
The zero point in the position axis is centered on the
star.  Negative offsets are to the east and positive to 
the west.  The dashed line marks the velocity of the
background H\,{\sc ii} region.  Only H$\alpha$ measurements are 
plotted; the [N\,{\sc ii}] lines are too weak for accurate 
measurements.}
\end{figure}

\section{Discussion}

\subsection{The formation of ring nebulae}

To determine the nature of the two ring nebulae around Sk$-$69 279 and 
Sk$-$69 271, it is worth  
looking into the mass loss history and how these winds interact with their 
environment to form ring nebulae. 

In the main-sequence phase
a fast stellar wind will sweep up the ambient medium to form a shell around  
the central star. 
These shells are called  {\it interstellar bubbles } (Weaver et al. 1977)
because they consist of mainly interstellar medium. An interstellar 
bubble blown by a 35\,M$_{\sun}$ in a medium 
of density  20 $\,{\rm cm^{-3}}$ can reach a typical radius of 38\,pc at
the end of the main-sequence phase (Garc{\'\i}a-Segura 
et al. 1996b).

The most massive stars, with an initial mass 
$M_{\rm ZAMS}$ $\geq$ 50 M$_{\odot}$, will lose half of their
mass during the main-sequence stage before evolving into 
Luminous Blue Variables (LBVs).
LBVs populate an area in the upper HR Diagram, called 
the Humphreys-Davidson limit (e.g. Humphreys \& Davidson 1979;  Humphreys \& 
Davidson 1994; Langer et al.\ 1994). 
At this point of their evolution the stars are very unstable
and have a very high mass loss rate around 10$^{-4}$\,M$_{\sun}\,
{\rm yr}^{-1}$. The strong stellar wind as well as the giant eruptions
during this phase  will strip a high amount of
mass from the star, preventing the star from reaching the red supergiant phase.
Since large amounts of mass have been ejected,  LBVs 
are often surrounded by small circumstellar nebulae (Nota et al.\ 1995;
Garc{\'\i}a-Segura et al. 1996a).

Less massive stars never reach this unstable phase; instead they  
evolve into  red 
supergiants after spending roughly  10$^6$ yr as main sequence O stars.
A 35\,M$_{\odot}$ star will lose about 2.5\,M$_{\odot}$ during its 
main-sequence phase with a wind velocity of about 1000\,km\,s$^{-1}$.
At the red supergiant phase, the 
wind becomes slower ($\simeq$ 20\,km\,s$^{-1}$) but much denser and a total 
amount of
18.6\,M$_{\odot}$ will be shed in the short (2.3 $\times$ 10$^5$ yr) time
before the star turns into a Wolf-Rayet star or blue
supergiant (Garc{\'\i}a-Segura et al. 1996b).

The wind of an  evolved star will contain processed
material. The dense wind at  the red supergiant phase or the LBV 
phase is enriched with CNO processed material. Consequently, a 
{\it circumstellar bubble}, formed by fast wind sweeping up the slow wind
or a LBV nebula  will show abundance anomaly. 

These possible evolutionary scenarios predict the formation of different types
of ring nebulae around massive stars. Furthermore, the physical properties of
the ring nebulae are tied in with the evolutionary states of the central
stars. By comparing the observational results of the ring nebulae of 
Sk$-$69 279 and Sk$-$69 271 with the predictions, we may determine the formation
mechanism of the ring nebulae and the evolutionary state of theses stars.  
 
\subsection{The nature of the nebula around Sk$-$69 279}

The morphology of the ring nebula around Sk$-$69 279 suggests that it 
could  be 
(1) an interstellar bubble blown by the star in the main sequence stage,
(2) a circumstellar bubble formed by the blue supergiant wind sweeping up the 
previous red supergiant wind, or
(3) a circumstellar bubble consisting of ejecta during an LBV phase of the 
star.

To distinguish among these possibilities we need to know the nitrogen 
abundance of the nebula.
If the nitrogen abundance in the nebula is similar to that of the ambient 
interstellar medium, the bubble is most likely an interstellar bubble.
If the nitrogen abundance is anomalous, it must contain processed stellar
material. Since the LBV ejecta is more nitrogen enriched, 
as much as 13 times the original abundance  (Garc{\'\i}a-Segura et al. 
1996a), than the RSG wind,
 which has 3 times the original abundance (Garc{\'\i}a-Segura et al. 
1996b),
the degree of nitrogen enrichment in the nebulae
can be used to differentiate between these two possibilities.
To diagnose the nitrogen abundance we use the
[N{\sc ii}]$\lambda$\,6583\AA/H$\alpha$ ratio extracted from the echelle data.
 
The [N{\sc ii}]$\lambda$\,6583\AA/H$\alpha$ ratio of the bubble 
is $\simeq$ 0.70  $\pm$ 0.02, while that of
the background is only $\simeq$ 0.07  $\pm$ 0.02. 
Assuming similar ionisation and excitation conditions in the bubble and
the ambient medium, a factor of 10 difference in the 
[N{\sc ii}]$\lambda$\,6583\AA/H$\alpha$ ratio implies a factor of 10 
enhancement in nitrogen abundance.  
Therefore it is likely that the bubble is nitrogen enhanced and contains
stellar material. 
This enhancement is too high for a red supergiant wind, therefore this 
bubble probably contains LBV ejecta.

For a LBV nebula (LBVN), the expansion velocity of Sk$-$69 279's ring, 
14\,km\,s$^{-1}$, is on the low end 
of the range  reported for other LBVNs (Nota et al. 1995). 
However, the size of Sk$-$69 279's ring, 4.5\,pc, is larger than the 
other known LBVNs. The dynamic time, defined as (radius)/(expansion velocity),
of Sk$-$69 279's ring is 1.5 $10^{5}$ yr, the largest among all known 
LBVNs. It is possible that the expansion has slowed down and
the dynamic age of the nebula is really lower.

The origin of the curious feature knot E is not clear.
It moves faster than the shell expansion, indicating that it belongs to a 
different kinematic system. Yet the [N{\sc ii}]$\lambda$\,6583\AA/H$\alpha$ 
ratio of the knot is similar to those in the shell.
It is possible that the knot originates from a  later and faster ejection and 
appears to be interacting with the shell.
It is also possible that the knot results from a fragmentation of the shell
and has been accelerated by stellar wind.
High resolution images and hydrodynamic modeling are needed to 
determine the nature of this knot.

\subsection{The nature of the nebula around Sk$-$69 271}

The morphology of the ring nebula around Sk$-$69 271, an arc, suggests a 
half-shell. Contrary to this impression the echelle spectra
shows a large expanding shell (radius $\sim$ 9\,pc) and the arc (radius 
$\sim$ 3\,pc) is only a small part on the approaching side of the shell.
To distinguish between the circumstellar and interstellar origin of the bubble,
again we use the  [N{\sc ii}]$\lambda$\,6583\AA/H$\alpha$ ratio.
We measured a ratio of 0.10 $\pm$ 0.02 in the expanding shell,
0.15 $\pm$ 0.02 in the arc, and 0.08 $\pm$ 0.02 in the ambient H\,{\sc ii}
region. 
These values do not argue for a nitrogen abundance enhancement in 
the ring nebula,
except possibly in the arc. Therefore, the shell consists mainly of
interstellar material and is an interstellar bubble.

We have associated the arc with the shell,
because it is very likely to be the  result of an interaction of the bubble
and an ambient interstellar filament or sheet. In such an interaction 
the projected shape of the interaction region will be dictated by
the geometry of shell. This kind of interstellar features
are probably common in this region at the outskirts of N\,160,
as the H$\alpha$ image (Fig.\,1) shows filamentary structures in the 
vicinity and some of the filaments are detected at different velocities in the
echelle data (Fig.\,3c). 

We can calculate the dynamic age of this interstellar bubble, $t_{\rm dyn}  =
\eta (\rm R/v_{\rm exp}$), where R is the radius of the shell and 
v$_{\rm exp}$ the expansion velocity.  
$\eta$ = 0.6 for an energy-conserving bubble (Weaver et al. 1977) or 
0.5 for a momentum-conserving bubble (Steigman et al. 1975). 
The bubble around Sk$-$69 271 would have a kinematic age of about 2 $10^5$yr.
Interestingly this age is smaller than the stars main-sequence life time.
This may imply complexities in the stellar mass loss history and in 
the density structure of the interstellar environment.  

Next we consider the ionisation of the nebula around Sk$-$69 271.
The interstellar bubble is not clearly recognizable in the H$\alpha$ 
image, probably due to the combined effects of low surface brightness of 
the bubble and confusion from the many foreground/background filaments in 
this region.     
We integrated the H$\alpha$ flux within a radius of 35\arcsec, the shell
radius determined from the echelle data.
The resulting H$\alpha$ flux, $9 \times 10^{34}$ erg s$^{-1}$, represents 
an upper limit for the shell, since it includes segments of possible 
foreground/background filaments. 
Converting the H$\alpha$ flux into the number of Lyman continuum 
photons necessary to ionize the gas, results in an upper limit of 
log N$_{Ly\alpha}$ $<$ 46.8.
We can compare this value with the model predictions (Panagia 1973), which 
give an ionizing flux of log N$_{Ly\alpha}$ = 46.18 for a B2 supergiant
as Sk$-$67 271. 
Since we only could derive an upper limit for the necessary photon 
flux, it is still possible, that Sk$-$69 271 is responsible for the ionisation 
of the interstellar bubble.

\section{Conclusion and summary}

In this paper we report on two stars which have been found to be surrounded
by ring nebulae. For Sk$-$69 279 we found a perfectly round ring structure 
and a  high [N {\sc ii}]$\lambda$\,6583\AA/H$\alpha$ ratio that  indicates 
processed material. The ratio is 10 times higher than the background. 
Therefore we suggest this ring nebula to be an older LBV ejecta.
No variability of the star or other signs of 
LBV activity have been reported, but the 
spectral type of Sk$-$69 279 is consistent with that of a quiescent LBV
(Wolf 1992, Shore 1993).

For Sk$-$69 271 we found an arc in the H$\alpha$ image, but our echelle 
spectroscopic observation reveals a larger expanding shell with the
arc being part of the approaching surface.
The [N {\sc ii}]$\lambda$\,6583\AA/H$\alpha$ ratio leads to the conclusion
that the shell is an interstellar bubble.

Deep, high-resolution images are needed to 
study the fine-scale structure of the circumstellar bubble around Sk$-$69 279 
and to reveal the morphology of the interstellar bubble around Sk$-$69 271.
Variability observations of Sk$-$69 279 are needed to verify its LBV nature.

\acknowledgements
{DJB thanks the Alexander von Humboldt 
Foundation for support through the Feodor Lynen Fellowship program.}

\end{document}